\def\confversion{0}
\def\ifconf{\ifnum\confversion=1}
\def\ifnotconf{\ifnum\confversion=0}

\documentclass[11 pt]{article}
\def\showauthornotes{1}
\def\showkeys{0}
\def\showdraftbox{0}

\usepackage{xspace,xcolor,enumerate}
\usepackage{amsmath,amssymb}
\usepackage{amsthm}
\usepackage{xfrac}
\usepackage[toc,page]{appendix}
\usepackage{thmtools}
\usepackage{thm-restate}
\usepackage{color,graphicx}
\usepackage{paralist}
\usepackage{thm-restate}

\ifnum\showkeys=1
\usepackage[color]{showkeys}
\fi

\definecolor{darkred}{rgb}{0.5,0,0}
\definecolor{darkgreen}{rgb}{0,0.35,0}
\definecolor{darkblue}{rgb}{0,0,0.55}

\usepackage[pdfstartview=FitH,pdfpagemode=UseNone,colorlinks,linkcolor=darkblue,filecolor=darkred,citecolor=darkgreen,urlcolor=darkred,pagebackref]{hyperref}

\usepackage[T1]{fontenc}
\usepackage{mathtools,dsfont,bbm}
\usepackage{mathpazo}

\usepackage[capitalise,nameinlink]{cleveref}
\ifnum\showauthornotes=1
\newcommand{\Authornote}[2]{{\sf\small\color{red}{[#1: #2]}}}
\newcommand{\Authorcomment}[2]{{\sf \small\color{gray}{[#1: #2]}}}
\newcommand{\Authorfnote}[2]{\footnote{\color{red}{#1: #2}}}
\else
\newcommand{\Authornote}[2]{}
\newcommand{\Authorcomment}[2]{}
\newcommand{\Authorfnote}[2]{}
\fi

\ifnum\showdraftbox=1
\newcommand{\draftbox}{\begin{center}
  \fbox{%
    \begin{minipage}{2in}%
      \begin{center}%
        \begin{Large}%
          \textsc{Working Draft}%
        \end{Large}\\
        Please do not distribute%
      \end{center}%
    \end{minipage}%
  }%
\end{center}
\vspace{0.2cm}}
\else
\newcommand{\draftbox}{}
\fi

\newtheorem{theorem}{Theorem}[section]

\newtheorem{lemma}[theorem]{Lemma}
\newtheorem{remark}[theorem]{Remark}
\newtheorem{proposition}[theorem]{Proposition}

\newtheorem{claim}[theorem]{Claim}

\crefname{nothing}{}{}
\crefformat{nothing}{(#2#1#3)}

\def\FullBox{\hbox{\vrule width 6pt height 6pt depth 0pt}}

\def\qed{\ifmmode\qquad\FullBox\else{\unskip\nobreak\hfil
\penalty50\hskip1em\null\nobreak\hfil\FullBox
\parfillskip=0pt\finalhyphendemerits=0\endgraf}\fi}

\def\qedsketch{\ifmmode\Box\else{\unskip\nobreak\hfil
\penalty50\hskip1em\null\nobreak\hfil$\Box$
\parfillskip=0pt\finalhyphendemerits=0\endgraf}\fi}

\def\to{\rightarrow}
\def\eps{\varepsilon}
\def\epsilon{\varepsilon}

\def\eps{\epsilon}

\def\phi{\varphi}
\def\cal{\mathcal}

\renewcommand{\bar}{\overline} 

\newcommand{\mper}{\,.}
\newcommand{\mcom}{\,,}

\newcommand{\R}{{\mathbb R}}

\newcommand{\N}{{\mathbb{N}}}

\newcommand{\zo}{\{0,1\}}

\newcommand{\gaussian}[2]{{\cal N(#1, #2)}}

\usepackage{nicefrac}
\makeatletter
\def\norm#1{%fancy norm command
  \@ifnextchar\bgroup%
   {\normalpnorm{#1}}%
   {\defaultnorm{#1}}%
}
\def\defaultnorm#1{%
    \renderNorm{#1}{}
}
\def\normalpnorm#1#2{%
   \@ifnextchar\bgroup%
   {\banaxnorm{#1}{#2}}%
   {\renderNorm{#2}{#1}}%\ensuremath{\left\lVert {#2} \right\rVert_{{#1}}}}%
}

\def\banaxnorm#1#2#3{%
    \renderNorm{#3}{#1\to#2 }
}

\def\renderNorm#1#2{%
    \@ifnextchar^%
    {\fixExponent{#1}{#2}}%
    {\ensuremath{\mathchoice%
        {\lVert #1 \rVert_{#2}}%
        {\lVert #1 \rVert_{#2}}%
        {\lVert #1 \rVert_{#2}}%
        {\lVert #1 \rVert_{#2}}}%
    }%
}
\def\fixExponent#1#2^#3{%
    \ensuremath{{\lVert #1 \rVert^{#3}_{#2}}}%
}

\def\enorm#1#2{%
   \@ifnextchar\bgroup%
   {\norm{L_{#1}}{L_{#2}}}%
   {\norm{L_{#1}}{#2}}%
}

\def\cnorm#1#2{%
   \@ifnextchar\bgroup%
   {\norm{\ell_{#1}}{\ell_{#2}}}%
   {\norm{\ell_{#1}}{#2}}%
}

\makeatother

\newcommand{\mydot}[2]{\ensuremath{\left\langle #1, #2 \right\rangle}}
\newcommand{\mysmalldot}[2]{\ensuremath{\langle #1, #2 \rangle}}

\newcommand{\one}{{\mathbf{1}}}

\newcommand{\Esymb}{\mathbb{E}}
\newcommand{\Psymb}{\mathbb{P}}

\makeatletter
\def\Pr#1{%
    \ProbabilityRender{\Psymb}{#1}%
}

\def\Ex#1{%
    \ProbabilityRender{\Esymb}{#1}%
}

\def\ProbabilityRender#1#2{%fancy probability command
  \@ifnextchar\bgroup%
  {\renderwithdist{#1}{#2}}
   {\singlervrender{#1}{#2}}
}
\def\singlervrender#1#2{%
   \ensuremath{\mathchoice
       {{#1}\left[ #2 \right]}
       {{#1}[ #2 ]}
       {{#1}[ #2 ]}
       {{#1}[ #2 ]}
   }
}
\def\renderwithdist#1#2#3{%
   \@ifnextchar\bgroup
   {\superfancyrender{#1}{#2}{#3}}
   {\ensuremath{\mathchoice
      {\underset{#2}{#1}\left[ #3 \right]}
      {{#1}_{#2}[ #3 ]}
      {{#1}_{#2}[ #3 ]}
      {{#1}_{#2}[ #3 ]}
     }
   }
}
\def\superfancyrender#1#2#3#4#5{
   \ensuremath{\mathchoice
      {\underset{#1}{{#1}}\left#4 #3 \right#5}
      {{#1}_{#2}#4 #3 #5}
      {{#1}_{#2}#4 #3 #5}
      {{#1}_{#2}#4 #3 #5}
   }
}
\makeatother

\newfont{\inhead}{eufm10 scaled\magstep1}

\newcommand{\calB}{{\cal B}}

\newcommand{\calP}{{\cal P}}

\newcommand{\id}{\mathrm{Id}}
\renewcommand{\bar}[1]{\ensuremath{\overline{#1}}}

\newcommand{\classfont}[1]{\textsf{#1}}

\newcommand{\NP}{\classfont{NP}\xspace}
\usepackage[colorinlistoftodos,textsize=tiny,textwidth=2cm,color=red!25!white,obeyFinal]{todonotes}



\newcommand{\SSS}{\mathbb{S}}

\newcommand{\bbE}{\mathbb{E}}

\newcommand{\bbY}{{\mathbb Y}}

\newcommand{\bb}[1]{{\mathbb #1}}

\usepackage[top=1in, bottom=1in, left=1in, right=1in]{geometry}

\crefformat{equation}{(#2#1#3)}
\Crefformat{equation}{(#2#1#3)}

\DeclareUnicodeCharacter{2113}{\ensuremath{\ell}}
\allowdisplaybreaks
\begin{document}

%\title{Approximate Polynomial Optimization Over the Unit Sphere}

\title{Inapproximability of ~$\ell_1\to \ell_2$~ Subspace Distortion}

\author{
Vijay Bhattiprolu\thanks{vbhattip@uwaterloo.ca} \\
University of Waterloo
}

% \date{April 2017}

\setcounter{page}{0}
%\date{\today}
\date{}

\maketitle
\draftbox
\thispagestyle{empty}

% \vspace{-15 pt}

We show, assuming NP is not contained in randomized quasipolynomial time, almost polynomial factor inapproximability for maximizing the $\ell_2/\ell_1$ ratio over a real subspace -- a problem also known as subspace distortion. 
This is a central problem in compressed sensing but also has connections to many areas such as real error correcting codes, dictionary learning, hypercontractive operator norms, small-set expansion, quantum information and random matrix theory. 

Distortion seems to be at the heart of building inapproximability theory for continuous optimization. 
Extending our results to the case of $\ell_2/\ell_p$ for all $p<2$ would lead to a near characterization of the convex sets over which one can efficiently and approximately maximize an arbitrary quadratic form. 

A recurring obstruction appears in hardness considerations of distortion -- how does one use a sparsity measure to test proximity to an exponential sized discrete set over the reals? 
This issue also appears in a host of other problems that involve measures of sparsity, shortness, density or expansion. 
We believe hardness of distortion is an intermediate step to understanding the complexity of small-set expansion and quantum separability, both of which are important directions of research. 

The main insight in our hardness reduction is the construction of a randomized affine linear map $T:\R^n\to \R^m$ such that the distortion of $Tx$ 
measures proximity to the hypercube $\{\pm 1\}^{n}$. Crucial to our construction is subsampling $x$ at *all* dyadic sample sizes, which gives an inverse polylogarithmic gap. This is a highly non-local tester, and thus departs from the local testing paradigm in PCP theory. 

The map $T$ we construct is also interesting from the standpoint of high dimensional geometry. It is a near-isometry, 
i.e., mapping $\SSS^{n-1}$ into a higher dimensional spherical shell, 
and moreover, $T$ sends the hypercube $\{\pm \sfrac{1}{\sqrt{n}}\}^{n}$ to points of distortion similar to a shifted-gaussian vector, and $T$ sends the points of $\SSS^{n-1}$ far from $\{\pm \sfrac{1}{\sqrt{n}}\}^{n}$ to points of super-gaussian distortion. Informally, $T$ inverts sparsity on $\SSS^{n-1}$. 

Our reduction is also quite succinct, and so it suggests a rather simple distribution of subspaces for which approximately finding a planted sparse vector is intractable.

\newpage

\pagenumbering{arabic}
\setcounter{page}{1}

% \phantomsection
% \addcontentsline{toc}{section}{References}

%%\mgnote{Test:
%%
%%\[
%%    \norm{2}{3}{A^T}^2
%%    \left\lVert\hat{A}^T\right\rVert_{2\to 3}^2
%%\]
%%
%%$
%%    \norm{2}{3}{A^T}
%%    \left\lVert\hat{A}^T\right\rVert_{2\to 3}^2
%%$
%%}

%$\norm{2}{3}{\hat{A}^T}$

%\section{Introduction}\label{sec:intro}
%\input{intro.tex}

\section{Introduction}
The $\ell_1\to\ell_2$ distortion $\Delta_{1\to 2}(V)$ of a subspace $V\subseteq \R^n$ is defined as 
\[
    \Delta_{1\to 2}(V) := \sup_{x\in V\setminus \{0\}} \Delta_{1\to 2}(x) \quad\text{where } 
    \Delta_{1\to 2}(x):= \sqrt{n}\cdot \norm{2}{x}/\norm{1}{x}\mper
\]
where $\norm{p}{x}:=(\sum_i |x_i|^p)^{1/p}$. For $1\leq p< 2$, we use $\Delta_{p\to 2}$ to denote the analogous quantity with $\Delta_{p\to 2}(x):= n^{1/p-1/2}\cdot\norm{2}{x}/\norm{p}{x}$. 

Any $x\in\R^n$ satisfies $\sqrt{n/\norm{0}{x}}\leq \Delta{1\to 2}{x} \leq \sqrt{n}$, where $\norm{0}{x}$ denotes the Hamming weight of $x$, i.e., the number of non-zero entries of $x$. 
The lowest value $\Delta_{1\to 2}(x)=1$ is attained iff $x$ is (upto scaling) a point in the hypercube 
$\{\pm 1\}^n$ and $\Delta_{1\to 2}(x)=\sqrt{n}$ iff $x$ is (upto scaling) an elementary vector. 
So $\Delta_{1\to 2}(x)$ is an analytic measure of sparsity of $x$, where $\Delta_{1\to 2}(x)$ increases 
as $x$ gets sparser. Distortion is a more robust measure of sparsity than the Hamming metric which is highly sensitive to small perturbations over the reals.

For a subspace $V$, $\Delta_{1\to 2}(V)$ is a real analogue of the minimum distance of an error correcting code, 
since $\Delta_{1\to 2}(V) =O(1)$ implies all nonzero vectors in $V$ are near-maximally spread. 
Such a subspace $V$ also has a geometric interpretation -- namely $V\cap \mathrm{Ball}(\ell_1^n)$ is nearly-euclidean (spherical). 
$\Delta_{1\to 2}(V)$ is an intensively studied quantity appearing in distinct areas. 
It has been studied in compressed sensing~\cite{guruswami_hsieh_raghavendra:SICOMP2026, jia_et_al:JSC2025,tao:SISC2022,zeng_yu_pong:SIOPT2021,rahimi_et_al:SISC2019,yin_esser_xin:CIS2014,hurley_rickard:TIT2009,hoyer:JMLR2004,KT07}, in the context of explicit error correcting codes over reals~\cite{guruswami2023sparsityellprestrictedisometry, guruswami2010almost, guruswami2008euclidean,Indyk06,indyk:STOC2007,lovett_sodin:CCM2008,linial_london_rabinovich:Combinatorica1995,ArtsteinAvidanMilman2006}, in the subexponential algorithm for small-set expansion~\cite{AroraBarakSteurer2015}, in dictionary learning ~\cite{berthelier_et_al:CVPRW2019,garg_et_al:IJCAI2017,barak_kelner_steurer:STOC2015}, in approximation theory and random matrix theory~\cite{Kashin97,garnaev_gluskin:DAN1984,FigielLindenstraussMilman1977,  guruswami_hsieh_raghavendra:SICOMP2026,guruswami_manohar_mosheiff:CCC2022,rudelson:CRMath2006}. Closely related variants are tightly connected to quantum separability~\cite{BBHKSZ12,HM13} and to polynomial maximization over convex sets~\cite{bhattiprolu2021framework,BBHKSZ12}. 
Despite intensive algorithmic effort, the best approximation ratio in both the worst and average cases remains $n^{\Omega(1)}$\cite{GuthMaldagueUrschel2025,Steinberg05,barak_kelner_steurer:STOC2015,DiakonikolasHopkinsPensiaTiegel2025,guruswami_hsieh_raghavendra:SICOMP2026}. 
 
Our primary motivation for studying $\Delta_{1\to 2}(V)$ is its centrality in the inapproximability theory 
of problems over the reals. Our main result is almost polynomial factor hardness for this problem assuming \NP can't be solved in randomized quasipolynomial time. This nearly settles its worst case approximability: 
\begin{restatable}[Almost Polynomial Factor Hardness of Subspace Distortion]{thm}{main}
\label{main}~\\
    Assume $\NP\not\in \cup_{d\in\N}\mathrm{BPTime}(n^{\log^{d} n})$. Then for any fixed $\eps>0$, no polynomial time algorithm can approximate 
    $\Delta_{1\to 2}(V)$ within a factor of $2^{\log^{1-\eps}n}$ on an input subspace 
    $V\subset \R^n$ (specified by a basis of its kernel).     
\end{restatable}

\begin{remark}
    As an immediate consequence, the same inapproximability result holds for 
    the following related problems:
    \begin{enumerate}[(a)]
        \item $\Delta_{p(n)\to 2}(V)$ for sufficiently slowly growing $p$, i.e., $p(n)\in [1,~1+1/\log^\eps n]$.

        \item  Maximizing a quadratic form over the $\ell_1^n$ ball (by considering 
        the reduction $V\mapsto ~\max_{\norm{1}{x}\leq 1} \norm{2}{\Pi_V x}^2-\beta\norm{2}{\Pi_{V^\perp}x}^2$ for a polynomially large non-negative scalar $\beta$). This is an intriguing contrast to the case of square-free quadratics, for which there is a PTAS (see for instance \cite{deKlerk08}). 

        \item $\ell_1$-low rank approximation via the reduction in \cite{BanEtAl2019}, for which 
        hardness was studied in multiple works~\cite{MahankaliWoodruff2021,ChierichettiEtAl2017,SongWoodruffZhong2017,GillisVavasis2018} and 
        was only known under stronger assumptions such as the small-set expansion hypothesis or ETH. 
        Our result also implies that assuming \NP cannot be solved in time $2^{n^{o(1)}}$, an exponential dependence on 
        the rank parameter $k$ is necessary. 
    \end{enumerate}
\end{remark}
\noindent
Inapproximability for $\Delta_{1\to 2}$ and closely related variants has been raised as a question in several  works~\cite{guruswami_hsieh_raghavendra:SICOMP2026, BBHKSZ12,BV11,HouSo2014, HillarLim2013,deKlerk08,deKlerk2008SimplexHypercubeSphere,KindlerNaorSchechtman2010, bhattiprolu2021framework,BGGLT23}. 
Previously only $1+1/n^{\Omega(1)}$ factor \NP-Hardness was known~\cite{tao:SISC2022}, and almost polynomial hardness was known under the small set expansion hypothesis~\cite{BBHKSZ12}. 

Many works are dedicated to the average case version of finding planted sparse vectors in a random subspace~\cite{guruswami_hsieh_raghavendra:SICOMP2026, DiakonikolasHopkinsPensiaTiegel2025,DingHua2023, HopkinsSchrammShiSteurer2016,BKS15}. A surprising feature of our reduction is that it also provides a relatively simple distribution over random subspaces for which there is evidence of strong hardness\footnote{we remark that it does not appear very difficult to derandomize our reduction given known results about samplers and euclidean sections. We chose to focus on the randomized reduction in this work for its simplicity.}. Assume hardness of finding a planted boolean vector in a random subspace $R$, which by now has several papers evidencing that it is hard to approximate within some fixed constant~\cite{MohantyRaghavendraXu2020,GhoshEtAl2020,MaoWein2025}. Our reduction suggests that if $T$ is a random affine linear map with sparse gaussian rows (at all dyadic sparsity scales), then it is hard to $(1+1/\log^{O(1)}n)$-approximately find a planted $\Delta_{1\to 2}$-sparse vector in $T(R)$. 

\subsection{Motivation}
Polynomial maximization over convex sets is a rich and expressive class of continuous optimization problems with a nuanced approximation landscape. $\Delta_{p\to 2}$ 
plays a central role in understanding the approximability of this landscape, and we elaborate on this below. 
\paragraph{Revealing the Approximability Landscape of Polynomial Maximization over Convex Sets.}
Any convex program may be rewritten as maximizing a linear function over a convex set, which covers a wide ranging list of optimization problems. 
When we consider the generalization of maximizing fixed degree polynomials over a convex set, 
the expressiveness increases overwhelmingly-- now capturing many problems (both continuous and discrete) that admit 
non-trivial approximation algorithms (such as maximum eigenvalue, max cut, grothendieck's inequality, non-commutative grothendieck's inequality, constraint satisfaction problems, robust PCA), but also many problems for which the best approximation known is $n^{\Omega(1)}$ (such as $\Delta_{p\to 2}$, sparse PCA, densest k-subgraph, densest subhypergraph). 

It was shown in \cite{bhattiprolu2021framework} that if \cref{main} were extended to all $p<2$, then one would obtain a near characterization of convex sets $C$ given via membership oracle, where one can approximate $\max_{x\in C} Q(x)$, given an 
arbitrary quadratic form $Q(x)$ as input. 
We discuss this in more detail below.

Let $(C_n\in \R^n)_{n\in \N}$ be a family of convex bodies of increasing dimension. 
\cite{bhattiprolu2021framework} defines the following body $\mathcal{U}(C_n)$ that encodes all gaussian random walks that stay within $C_n$ in expectation: 
\[
    \mathcal{U}(C_n):= 
    \{\Sigma\in \R^{n\times n}~|~ \Sigma\succeq 0,~\Ex{g\sim\Sigma}{\norm{C_n}{g}^2}\leq 1\}, 
\]
where $g$ is a gaussian vector of covariance $\Sigma$, and $\norm{C_n}{\cdot}$ 
denotes the minkowki functional of $C_n$. 
Finally let $\mathrm{Convexity}(\mathcal{U}(C_n))$ denote the smallest possible distortion with which $\mathcal{U}(C_n)$ can be approximated by a convex body.\footnote{$\mathrm{Convexity}(\mathcal{U}(C_n))$ equals the type-2 constant of $C_n$, which is usually defined differently, and is a well studied notion in the convex geometry literature. It is closely related to the smallest distortion with which $C_n$ can be approximated by a uniformly smooth convex body.} 

The following algorithmic dichotomy was shown in \cite{bhattiprolu2021framework}: 
\begin{enumerate}[-]
    \item Assuming for all $p\in [1,2)$ that $\Delta_{p\to 2}$ cannot be 
    $2^{\log^{1-\eps}n}$-approximated in quasipolynomial time, 
    then whenever $\mathrm{Convexity}(\mathcal{U}(C_n))$ grows too quickly, specifically $\mathrm{Convexity}(\mathcal{U}(C_n))=n^{\Omega(1)}$, the same 
    almost-polynomial factor inapproximability holds for $\max_{x\in C_n} Q(x)$. 

    \item On the other hand if $\mathrm{Convexity}(\mathcal{U}(C_n)) = O(1)$ and if additionally $\mathcal{U}(C_n)$ admits an $O(1)$-approximate separation oracle, 
    there is a $\mathrm{poly}(n)$ time algorithm for that approximates quadratic maximization over $(C_n)$ within a constant factor. 
\end{enumerate}
One can view the above as a new incarnation of an important theme in 
the field of approximation algorithms -- specifically that convex relaxations predict 
approximability. The most well known example of this is the theorem of \cite{Raghavendra2008OptimalAA}, which establishes that assuming the unique games conjecture\footnote{for which there is a recent claimed proof~\cite{openai2026uniquegames}}, 
for any constraint satisfaction problem of constant locality, 
a canonical semidefinite programming relaxation of the problem obtains the optimal approximation factor among all polynomial time algorithms.

%$\Delta_{p\to 2}$ (as well as related $p\to 2$ operator norms) seems to play a central role in the inapproximability of problems over the reals. 
% This is a very expressive class of problems (indeed the linear case captures all of convex optimization) that subsumes many non-trivial approximable and non-approximable examples of continuous and discrete optimization problems. 
\paragraph{Small-Set Expansion}
$\Delta_{1\to 2}$ (and the closely related $p\to 2$ operator norms for $p<2$), is closely connected to small-set expansion~\cite{AroraBarakSteurer2015,BS14} (a central problem in inapproximability theory). For instance the ``Local Cheeger Theorem'' in \cite{BarakSteurerLocalCheeger} implies that if $L$ is the Laplacian of a graph $G$, then 
$\max_{\norm{1}{x}\leq \sqrt{\delta\cdot n}} \norm{2}{x}^2-x^TLx$ approximates the small-set expansion of $G$ for sets of size $\leq \delta\cdot n$. 
Thus \cref{main} can be viewed as an intermediate step towards the Small-set expansion hypothesis, since it establishes hardness for general quadratics, whereas the above instances are a special class of quadratics, arising from projectors to the top eigenspace of a graph. 

The ``Structure Theorem'' in \cite{BarakSteurerLocalCheeger} also shows how the 
celebrated subexponential time algorithm for small-set expansion due to \cite{AroraBarakSteurer2015} proceeds via reduction to $\Delta_{1\to 2}$. 
Unless $\NP$ can be solved in time $2^{n^{o(1)}}$, \cref{main} implies that $\Delta_{1\to 2}$ cannot be approximated in time $2^{n^{o(1)}}$.

%that the distortion of the top eigenspace of a graph approximates the Small-Set Expansion problem, whose complexity is of significant interest due to its connection with the Unique Games Conjecture. 

\paragraph{Quantum Separability.}
In \cite{HM13,BBHKSZ12} it was shown that the classical complexity of approximating the $p\to 2$ norm of a linear operator for $p<2$ (or equivalently, the $2\to q$ norm for $q>2$ by passing to the adjoint operator) is closely related to the complexity of quantum separability ("QMA vs. NEXP"); a central question in the field of quantum information. Recall $\norm{p}{2}{A}:= \max_{x\neq 0} \norm{2}{Ax}/\norm{p}{x}$. We believe \cref{main} makes progress towards inapproximability of $\norm{p}{2}{\cdot}$ and of quantum separability, as it implies hardness for the following generalization: $\max_{x\neq 0} \norm{2}{Ax}/\norm{p}{Bx}$, where $A,B$ are linear operators and $p\in[1,1/\log^\eps n]$. 

In fact, a key sub-component of our reduction extends to all $p\in [1,2)$ (namely, we have a logarithmic sized list of tests, such that for any sparse vector all of whose nonzero entries have equal magnitude, it is penalized by some test in the list). Hardness for $\norm{p}{2}{A}$ appears to require additional ideas beyond just inapproximability of $\Delta_{p\to 2}$, even though the problems are closely related.

\paragraph{Towards a Theory of Inapproximability over the Reals.}
Inapproximability theory is quite well fleshed out for discrete optimization (particularly optimization problems with local constraints), but the theory has stalled over the reals due to serious challenges. We believe the 
study of inapproximability of $\Delta_{p\to 2}$ could shed light on a host of other problems that pose 
similar barriers to existing techniques -- any problem where 
sparse solutions can have high objective value such as polynomial maximization over $\ell_p$ balls for $p\leq 2$,~ $p\to q$ operator norms for $p\leq q$, sparse PCA, graph expansion problems, densest k-subgraph, sparse recovery in compressed sensing, minimum distance codeword in a code (particularly large fields or fields of characteristic zero), shortest vector problems in integer lattices,  etc. 

Below we elaborate on the ``testing for sparse solutions'' obstruction in inapproximability theory. Our proof of \cref{main} provides a fairly general approach for detecting sparse solutions -- we subsample a solution $x\in\R^n$ at all dyadic scales, i.e. samples have sizes $\{2^{j}~:~j\in \log_2 n \}$.

\paragraph{Incompatibility of Sparse Solutions with the Local Testing Framework.}
We briefly describe the difficulty of excluding sparse solutions using local tests; local testing being the primary driver of a large fraction of inapproximability results. In contrast to the case of finite fields, not much is known about robustly locally testable real codes, i.e., codes where points $\eps$-far from the code in $\ell_2$-norm are rejected with constant probability by a local tester. There are even provable barriers under appropriate assumptions. Indeed in \cite{guruswami2023sparsityellprestrictedisometry}, it is shown that for any matrix whose kernel satisfies $\Delta_{1\to 2} = O(1/\sqrt{n})$ and has constant rate, 
$n^{\Omega(1)}$ of its rows must have sparsity $\omega(\log n)$. These types of barriers suggest the local gadget reduction paradigm from PCP theory is not sufficient. 

\emph{The key new contribution~(\cref{inversion:map}) in our reduction is a tester of the form $\Delta_{1\to 2}(Tx)$ for proximity to $\{\pm 1\}^n$ in $\ell_2$ distance, where $T$ is an affine linear map. $T$ performs highly non-local linear measurements of $x$ at polylogarithmically many locality scales, i.e., we query random subsets of $x$ of sizes $\{2^j:j\in [\log_2 n]\}$ (each with equal probability), which range from constant locality all the way to locality $n$.}

\section{Related Work: Error Correcting Codes, Lattices and Locally Dense Gadgets}
\label{related}
There is a rich and successful theory of inapproximability for the minimum distance of an error correcting code over finite fields and for the shortest vector in an integer lattice \cite{ajtai1998shortest,micciancio2001shortest, khot2005hardness, micciancio2012inapproximability,M14,  AK14, bhattiprolu25,wan2026nphardnesssvpeuclideanspace,hecht2025deterministic,hair2025svp}. These 
works use coding gadgets to overcome the sparse solutions issue discussed earlier. With the exception of a small number of recent works (\cite{bhattiprolu25,hecht2025deterministic,hair2025svp,BGGLT23}), all prior hardness results for error correcting codes and lattices made use of locally dense gadgets (see \cite{M14} for a detailed discussion), which are codes/lattices that have exponentially many codewords/lattice-points in a ball of radius strictly smaller than the minimum-distance/shortest-vector.

\textbf{Coding Challenges over Reals.} It seems quite challenging to devise locally dense subspaces over the reals as the understanding of real error correcting codes is lacking compared to finite fields. For instance, even though random subspaces over the reals do not have local density (they admit a weaker version for a radius above the minimum distance), the radius for which they are locally dense is only known upto constants. Many locally dense gadget constructions over finite fields involve  codes that have better rate vs. distance tradeoffs than random subspaces. However, current explicit constructions of codes over the reals 
have much worse parameters than random subspaces -- explicit construction of a real subspace with constant distortion and constant rate (which also implies a weak version of local density) is a long standing open problem~\cite{guruswami2008euclidean,guruswami2010almost,indyk:STOC2007,Indyk06,lovett_sodin:CCM2008,linial_london_rabinovich:Combinatorica1995}.

Even the exceptions of \cite{hair2025svp,hecht2025deterministic} that make use of coding gadgets related to Hadamard matrices, do not work for subspace distortion  
as they contain sparse vectors for the aforementioned reasons (in fact sparse solutions are discussed as an obstruction towards hardness of shortest $\ell_2$ vector in a lattice in \cite{hecht2025deterministic}). 

\paragraph{Randomized Coding Gadgets}
Random subspaces only satisfy a weak version of local density, however, in \cite{bhattiprolu25} it is shown that they have additional beneficial properties that allow them to be used in the inapproximability of minimum hamming distance (and in \cite{BGGLT23} for hardness of $\Delta_{p\to q}$ when $p\leq q< 2$) of a subspace over the reals (or other fields). \cite{bhattiprolu25} uses a random rademacher kernel as a coding gadget, and \cite{BGGLT23} uses a random gaussian subspace as a gadget\footnote{specifically, it uses a classical result of Dvoretzky and Milman on the embedding of $\ell_2^m$ inside $\ell_p^N$}.  

 \paragraph{Hypercube Proximity Testing and a Sparsity Inversion Map.}
Building on \cite{bhattiprolu25,BGGLT23}, we also construct a randomized gadget and use it to show inapproximability of $\Delta_{1\to 2}$. 
However, $\Delta_{1\to 2}$ behaves very differently from hamming weight and from $\norm{1+\eps}{2-\eps}{A}$, so neither Rademacher kernels nor gaussian 
random subspaces have the desired properties. As mentioned earlier, our gadget~\cref{inversion:map} is a tester of the form $\Delta_{1\to 2}(Tx)$ for proximity to $\{\pm 1\}^n$ in $\ell_2$ distance. $T$ is an affine random linear map, having sparse gaussian entries for a sparsity parameter that is selected randomly from a dyadic grid. 

The map in \cref{inversion:map} has an intriguing geometric interpretation. It is a near-isometric affine map (i.e. mapping the sphere $\SSS^{n-1}$ to a thin spherical shell in higher dimension), that moreover sends $\{\pm \sfrac{1}{\sqrt{n}}\}^n$ to somewhat sparse points 
(having distortion similar to that of a shifted gaussian random vector) 
and sends $\{x\in\SSS^{n-1}~:~ \mathrm{dist}_2(x,\{\pm \sfrac{1}{\sqrt{n}}\}^n)> \eps\}$ to more spread points (having $1+\eps/\log^{O(1)}n$ factor higher distortion than that of a shifted gaussian random vector). Informally, it inverts sparsity of points on $\SSS^{n-1}$.

\subsection{AI methodology}

\begin{flushleft}
  \renewcommand{\arraystretch}{1.3}
  \begin{tabular}{@{}l@{\qquad}c@{}}
    \textbf{Activity} & \textbf{AI used} \\
    Asking the research question & No \\
    Coming up with the approach & No \\
    Proof development & No \\
    Writing and exposition & No \\
    Checking for bugs & Yes \\
    Other supporting tasks & Locating references \\
  \end{tabular}
\end{flushleft}

\paragraph{Details}
Use of AI was minimal in this work. Prior to any interaction with AI, we obtained a proof of our main theorem for the special case of vectors whose nonzero entries have equal magnitude. 
We then tested our intuition on other vectors by asking Gemini and ChatGPT 5.5 to compute expectations of the form $\Ex{|1+\mysmalldot{\xi_d}{x}|}$ for a few test vectors (both gave similar answers). We then proved all claims from scratch without AI reference. Nothing in the reduction or proof changed materially from computing the expectations.

\section{Proof Overview}
\label{overview}

\subsection{Notational Conventions.}
$\R^n$ denotes $n$-dimensional Euclidean space and we index points/vectors $x\in \R^n$ using subscripts $x_i$. We use $\one$ to denote the all-ones whose dimension will always be finite but vary based on context (the dimension will always be clear from context). $x\oplus y$ denotes vector concatenation. $\SSS^{n-1}\subset \R^n$ denotes 
the unit sphere in $n$ dimensions.

For any $p\in [1,\infty)$, the (counting) $\ell_p$ norm is defined as $\norm{p}{x}:= (\sum_{i\in [n]}|x_i|^p)^{1/p}$. 
For $p=\infty$, we define as usual $\norm{\infty}{x}:= \max_{i\in [n]}\{|x_i|\}$. 
In several places it is more convenient to use expectation $L_p$ norms. For $L_p$ integrable functions $f:\R\to\R$, we use $\norm{L_p(\R)}{f}$ to denote $(\int_\R |f(x)|^p\,dx)^{1/p}$. 
For $x\in \R^n$, We use $\norm{L_p([n])}{x}$ to denote $\Ex{I\sim [n]}{|x_I|^p}^{1/p}$, where $I$ is sampled uniformly from $[n]$. 
Note that $\Delta_{1\to 2}(x) = \norm{L_2([n])}{x}/\norm{L_1([n])}{x}$, which appears a more natural relation than the scaled counting norm definition. 

We fix notation for the distance between sets $S,S'\subset \R^n$, $\mathrm{dist}_2(S,S') := \inf_{(x,x')\in S\times S'} \norm{2}{x-x'}$ to denote the $\ell_2$ distance from $S'$ to $S$.  

Throughout this document, $\tau:=\sqrt{(160/\eta)\log\log n}$, where $\eta\in (0,1)$ is an unspecified constant that only needs to be chosen sufficiently small. 
$\Phi(t)$ denotes the tail weight of a standard gaussian, i.e., $\Phi(t):= \Pr{|g|\geq t}$ for a standard gaussian $g$. 

\iffalse
%\section{Technical Overview.}
\subsection{Hypercube Hardness}
It is convenient for us to reduce from hardness of quadratic maximization over the hypercube. This choice is by no means special -- other starting points such as set cover can be made to work with similar ideas. 

We use the following result: 
\begin{theorem}[$\ell_2$-Robust Hardness of $\infty\to 2$ Operator Norm~\cite{BRS15}]
\label{infty:2}
    For any fixed $\eps,\eta,\delta>0$ sufficiently small, 
    given as input an $m\times n$ matrix $A$ satisfying $\norm{2}{2}{A}\leq 1$, it is NP-Hard to distinguish between the following cases: 
    \begin{compactenum}
        \item[(Completeness):] There exists $x\in\{\pm \frac{1}{\sqrt{n}}\}^n$ satisfying 
        \[\norm{2}{Ax}=1.\] 
        
        \item[(Soundness):] For any $x\in\R^n$ satisfying $\norm{2}{x}\leq 1$ and $\mathrm{dist}_2(x, (1+\eta)\mathrm{Ball}(\ell_\infty^n))\leq \eps_\infty$, 
        we have {\color{blue}
        \[\norm{2}{Ax}\leq (1+\eta)(\sqrt{2/\pi}+\eps)+\eps_\infty\mper\]}
    \end{compactenum}
\end{theorem}
\fi

\subsection{Testing $\ell_2$-Proximity to Hypercube}
The main goal of this section is to give a proof sketch of \cref{inversion:map}, which forms the backbone of our hardness proof. Informally speaking, it 
shows how we can test if a given vector $x\in \SSS^{n-1}$ is close to 
$\{\pm\sfrac{1}{\sqrt{n}}\}^n$ using a test of the form 
\[
    \norm{1}{T_{inv} \,x}\leq 1
\]
where $T_{inv}:\R^n\to \R^m$ is an affine linear map. 
We first design a test of the form 
\[
    \norm{L_1(\R)}{\Xi(x)}\leq 1
\]
where $\Xi$ is an affine linear map from $\R^n$ to probability distributions 
over $\R$. As is now routine in high dimensional geometry, one can then leverage the concentration of measure phenomenon to approximate $\Xi$ by a map $T_{inv}:\R^n\to \R^m$, where  $m=n^{O(1)}$. We do that in \cref{inv:sec}. 

\paragraph{Affine $\ell_1^n$ test for vectors that are spread.}
Observe that $(|a+b|+|a-b|)/2 = \max\{|a|,|b|\}$ which can be interpreted geometrically as the fact that $\ell_\infty^2$ embeds isometrically inside $\ell_1^2$, 
and consider the following ``Threshold-$\ell_1$-Test'': 
\begin{equation}
\label{thresh}
    \frac{1}{2}\sum_i |1+\sqrt{n}\cdot x_i|+|1-\sqrt{n}\cdot x_i| ~~=~~ 
    \max\{1,\sqrt{n}\cdot |x_1|\}+\dots +\max\{1,\sqrt{n}\cdot |x_n|\}
\end{equation}
Clearly the above quantity is minimized for $x\in \{\pm 1/\sqrt{n}\}^n$, 
and moreover the value is a constant factor larger, if $x$ has $\Omega(n)$ entries 
of magnitude $(1+\Theta(1))/\sqrt{n}$. Importantly, the Threshold-$\ell_1$-Test simultaneously penalizes high magnitude entries $|x_i|>1/\sqrt{n}$ while not rewarding 
low magnitude entries $|x_i|<1/\sqrt{n}$. 

\paragraph{Limitation of Threshold-$\ell_1$-Test.} 
This test does not yield a gap if $x$ is very sparse. Indeed consider $x:=e_1$ and 
note that the value is $\sqrt{n}+1-1/\sqrt{n}$ which is larger than the value of 
any point in $\{\pm 1/\sqrt{n}\}^n$, but only by a factor of $1+\Theta(1/\sqrt{n})$. 
The issue here is that we would like to use the $\ell_1$ norm to measure the Threshold-$\ell_2$ norm of $x$ (at threshold $1/\sqrt{n}$), and the Threshold-$\ell_1$ norm is only a good proxy for this when $x$ is highly non-sparse. 

\paragraph{Subsampling-$L_1(\ell_2^{n/2})$-Test for sparse vectors.}
In response to the above limitation, we consider the following (non-$\ell_1$) test that gives a constant separation between $e_i$ and  $\{\pm 1/\sqrt{n}\}^n$ (we will later embed this test into $\ell_1$). 

Let $I\sim [n]$ be a uniformly random element from $[n]$, and let $I_1,I_2,\dots I_{n/2}$ be i.i.d. copies of $I$. We refer to the following test as 
\begin{align}
\label{subsample:test}
    &\frac{1}{2}\Ex{|1+\sqrt{2}(x_{I_1}^2+\dots +x_{I_{n/2}}^2)^{1/2}|+|1-\sqrt{2}(x_{I_1}^2+\dots +x_{I_{n/2}}^2)^{1/2}|} \nonumber\\
    ~~=~~
    &\Ex{\max\{1,\sqrt{2}(x_{I_1}^2+\dots +x_{I_{n/2}}^2)^{1/2}\}}
\end{align}
Clearly any point $x\in \{\pm 1/\sqrt{n}\}^n$ obtains value $1$ 
since $2(x_{I_1}^2+\dots +x_{I_{n/2}}^2) = 1$ always. On the other hand, any 
elementary vector $x=e_i$ causes a visible spike, i.e., 
$2(x_{I_1}^2+\dots +x_{I_{n/2}}^2) = 2$ with probability 
$(1-1/n)^{n/2} = \Omega(1)$. In the remaining possibility where $2(x_{I_1}^2+\dots +x_{I_{n/2}}^2) = 0$, there is no reward since $\max\{1,\sqrt{2}(x_{I_1}^2+\dots +x_{I_{n/2}}^2)^{1/2}\}\geq 1$. Thus elementary vectors admit a gap of $1+\Omega(1)$. 

\paragraph{Combining tests by averaging.}
We have a new issue with the Subsampling-$L_1(\ell_2^{n/2})$-Test (\cref{subsample:test}) -- vectors such as $x:=(\sqrt{2/n},\dots \sqrt{2/n},0,\dots ,0)$ that were covered by the Threshold-$\ell_1$-Test, now 
only admit a gap of $1+O(1/\sqrt{n})$ on the Subsampling-$L_1(\ell_2^{n/2})$-Test since for such vectors, $2(x_{I_1}^2+\dots +x_{I_{n/2}}^2)$ is concentrated around $1$ with high probability.

A saving grace is that $\{\pm 1/\sqrt{n}\}^n$ is a common minimizer for both the 
Threshold-$\ell_1$-test (\cref{thresh}) and the Subsampling-$L_1(\ell_2^{n/2})$-Test (\cref{subsample:test}), which enables the use of an average  
of both tests:
\begin{equation}
\label{2:composite}
    \frac{1}{2}\Ex{\max\{1,\sqrt{2}(x_{I_1}^2+\dots +x_{I_{n/2}}^2)^{1/2}\}}
    ~+~\frac{1}{2}\max\{1,\sqrt{n}\cdot |x_1|\}+\dots +\max\{1,\sqrt{n}\cdot |x_n|\}
\end{equation}
It's not hard to verify that this composite test retains a gap of $1+\delta/2$ 
for any vector $x$ that has value at least $(1+\delta)$ times the value of $\{\pm 1/\sqrt{n}\}^n$ for at least one of \cref{thresh} and \cref{subsample:test}. 
It follows that \cref{2:composite} separates (with gap $1+\Omega(1)$), 
$\{\pm 1/\sqrt{n}\}^n$ from both elementary vectors as well as vectors of effective 
support $\Omega(n)$ that are far from $\{\pm 1/\sqrt{n}\}^n$. 

\paragraph{$L_1(\ell_2^d)$ Testing at all dyadic scales $d$.}
We are not yet done since vectors of intermediate support size, say $n^\delta$, 
do not yield a gap on either the Subsampling-$L_1(\ell_2^{n/2})$-Test or the Threshold-$\ell_1$-Test. We fix this by introducing few (logarithmically many) interpolates between these two tests at dyadic scales. With this new goal in mind, it is convenient to reinterpret \cref{2:composite} as combining a Subsampling-$L_1(\ell_2^{n/2})$-Test (sample size $n/2$) with a Subsampling-$L_1(\ell_2^{1})$-Test (sample size $1$). 
We define the following ``Multiscale-Subsampling-Test'': 
\[
    \Ex{J\sim [\log_2 n]}{\Ex{I_1,\dots ,I_{2^J}}{\max\{1,\sqrt{n/2^j}\cdot (x_{I_1}^2+\dots +x_{I_{2^j}}^2)^{1/2}\}}}
\]
where $J$ is sampled from $[\log_2 n]$ uniformly. 
It is not hard to check that for any $x\in\SSS^{n-1}$ that has threshold $\ell_2^2$-mass at least $\eps_\infty$ above the threshold $\sqrt{2/n}$, then there exists $j\geq 1$
such that $x$ has $\ell_2^2$-mass at least $\eps_\infty/\log_2 n$, restricted to the level set $\{i~:~ x_i^2 \in [2^j,2^{j+1})\}$. Furthermore, for this $j$ we have 
\begin{equation}
\label{fixed:scale}
    \Ex{I_1,\dots ,I_{2^{j-1}}}{\max\{1,\sqrt{n/2^j}\cdot (x_{I_1}^2+\dots +x_{I_{2^j}}^2)^{1/2}\}} = 1+\Omega(\eps_\infty/\log n)
\end{equation}
which implies that $x$ has value at least $1+\Omega(1/\log^2 n)$ on the 
Multiscale-Subsampling-Test, thereby giving a separation of $1+\Omega(1/\log^2 n)$ 
from $\{\pm\sfrac{ 1}{\sqrt{n}}\}^n$ (which has value $1$). We emphasize again here 
that $\{\pm\sfrac{ 1}{\sqrt{n}}\}^n$ being the common minimizer of \cref{fixed:scale} for all $j$, is crucial for the average of the tests to retain a gap. 

We are left with the embedding this test into $L_1(\R)$. 

\paragraph{Gap preserving embedding of $L_1(\ell_2^d)$ test into $L_1(\R)$.}
We begin by recalling a classical observation -- $\ell_2^n$ embeds isometrically inside $L_1(\R)$ using rotation invariance of the gaussian distribution: 
\[
    \Ex{|\mysmalldot{(g_1,\dots ,g_n)}{x}|} = \sqrt{2/\pi}\cdot \norm{2}{x}
\]
where $g_1,\dots ,g_n$ are i.i.d. standard gaussians. 

It is natural to combine this idea with the Multiscale-Subsampling-Test,  
yielding what follows. 

\paragraph{The Final $L_1(\R)$-Test}
Let $\bar{n}:=(4\log_e n)\,n$. 
Fix a parameter $\eta\in (0,1)$ and define $B:=\{1,\dots ,\lfloor\log_{1+\eta} \bar{n}\rfloor\}$\footnote{we bucket by powers of $1+\eta$ instead of dyadically to handle the case of $x$ having $\ell_2$-mass in the interval $[(1+\eta),(1+\eta)^2)/\sqrt{n}$ } and let $\Omega := \{(1+\eta)^j~:~j\in B\}$. 
For $d\in \N$, let $\xi_d\in\R^n$ be the vector random variable given by 
$\sqrt{n/d}(g_1 e_{I_1}+\dots +g_d e_{I_d})$. Let $\tau:=\sqrt{(160/\eta)\cdot \log \log n}$ be a scaling parameter, set so that $\Phi(\tau)=\widetilde{O}(\log^{80/\eta}n)$. Define $\xi$ to be the random vector $\xi_D$ where $D$ is chosen 
uniformly from $\Omega$. 
We are finally ready to define our Hypercube proximity test. 
\medskip 

\noindent
\emph{Multiscale $L_1(\R)$ Test of $\ell_2$-proximity to the Hypercube: $\norm{L_1(\R)}{1+\mysmalldot{\xi}{x}/\tau}$}
Observe that
    \begin{align}
        \norm{L_1(\R)}{1+\mysmalldot{\xi}{x}/\tau} &= \Ex{|1+\mysmalldot{\xi}{x}/\tau|} 
        \nonumber \\
        &= (1/2)(\Ex{|1+\mysmalldot{\xi}{x}/\tau|}+\Ex{|1-\mysmalldot{\xi}{x}/\tau|})
        &&(\text{since } \xi\equiv -\xi)
        \nonumber \\
        &= \Ex{\max\{1,|\mysmalldot{\xi}{x}|/\tau\}} \mcom
    \end{align}
a fact that we leverage throughout this work. 

We proceed to analyse this test.

\section{Randomized Construction of Sparsity Inversion Map}
We begin by showing a separation in expectation (which corresponds to an infinite dimensional map). 
In \cref{inv:sec}, we leverage the concentration of measure phenomenon and take an average over polynomially many samples to obtain a polynomial dimensional map. 
\begin{theorem}[Distortion Tester for Hypercube Proximity]
\label{bdd:test}
    There are sufficiently large universal constants $C,C'>1$ such that the following 
    holds. Fix any $\eta,\eps_\infty(0,1)$ sufficiently small. Then, 
    \begin{compactenum}
        \item[($\{\pm\sfrac{1}{\sqrt{n}}\}^n$ has a Sparse Image):] 
        For any $x\in \mathrm{Ball}(\ell^n_\infty)$,  
        \[\norm{L_1(\R)}{1+\mysmalldot{\xi}{x}/\tau}\leq \norm{L_1([\R])}{1+g/\tau} =1+O(\Phi(\tau))\]
        
        \item[(Sparse Points have a Spread Image):] $\forall~x\in\R^n$ such that  
        $\mathrm{dist}_2(x,\sqrt{(1+\eta)/n}\cdot \mathrm{Ball}(\ell^n_\infty))\geq \sqrt{\eps_\infty}\mcom$ 
        \[
            \norm{L_1(\R)}{1+\mysmalldot{\xi}{x}/\tau}~\geq ~ 
            1+\Omega(\eps_{\infty}\!\cdot\Phi(\tau)^{1-\eta/5})\mper
        \] 
    \end{compactenum}
    where $\Phi(\tau)= \widetilde{O}( 1/\log^{80/\eta}n)$. 
\end{theorem}

\begin{proof}~\\
    \textbf{\large $\{\pm\sfrac{1}{\sqrt{n}}\}^n$ has a Sparse Image:}~\\
    Consider any $x\in \frac{1}{\sqrt{n}}\mathrm{Ball}(\ell^n_\infty)$. 
    For any $d\in \N$, $\mysmalldot{\xi_d}{x}$ is distributed according to a gaussian 
    with mean $0$ and variance at most $1$, since $(n/d)(x_{I_1}^2+\dots +x_{I_d}^2) \leq 1$ with probability $1$. Thus $\mysmalldot{\xi}{x}$ is distributed according to a gaussian with mean $0$ and variance at most $1$.  
    So we have, 
    \[
        \Ex{\max\{1,|\mysmalldot{\xi}{x}|/\tau\}} = 
        1+O\Big(\frac{1}{\tau}\int_{\tau}^{\infty} \frac{t\,dt}{e^{-t^2/2}}\Big)
        = 
        1+O(\Phi(\tau))\mper
    \]

\noindent
    \textbf{\large Sparse Points have a Spread Image:}~\\
    Fix any $x'$ satisfying 
    $\mathrm{dist}_2(x',\sqrt{(1+\eta)/n}\cdot \mathrm{Ball}(\ell^n_\infty))\geq \sqrt{\eps_\infty}$. 
    \paragraph{Scaling monotonicity.}
    Observe that for any $\lambda\geq 1$, 
    \[\max\{1,|\mysmalldot{\xi}{x}|/\tau\}\leq 
    \max\{1,\lambda\cdot |\mysmalldot{\xi}{x}|/\tau\}\leq 
    \max\{1,|\mysmalldot{\xi}{\lambda\cdot x}|/\tau\}\mper \]  
    Thus $\Ex{\max\{1,|\mysmalldot{\xi}{x}|/\tau\}}\leq \Ex{\max\{1,|\mysmalldot{\xi}{\lambda\cdot x}|/\tau\}}$. 

    \paragraph{Suffices to consider points in constant radius $\ell_2$-ball.}
    If $\norm{2}{x'}\geq 3$, then set $x:= 3\cdot x'/\norm{2}{x'}$, so that $\norm{2}{x}=3$, otherwise set $x:=x'$. Note that $x$ satisfies $\mathrm{dist}_2(x,\sqrt{(1+\eta)/n}\cdot \mathrm{Ball}(\ell^n_\infty))\geq \sqrt{\eps_\infty}$. 
    Due to monotonicity of scaling observed above, it suffices to show that 
    \[
        \Ex{\max\{1,|\mysmalldot{\xi}{x}|/\tau\}}\geq 
        1+\Omega\Big(\frac{\eps_\infty}{\log^{8/\eta -3/2} n}\Big)
    \] 

    \paragraph{$(1+\eta)$-adic Bucketing of Entry Magnitudes.}
    We organize the entry magnitudes of $x$ by powers of $1+\eta$. 
    Let $\calB_{a}$ denote the set of all indices 
    $i$ such that $x_i^2 \in [a,(1+\eta)a)$. Let $x_{\calB_a}$ denote 
    the restriction of $x$ to the indices in $\calB_a$. 

    \paragraph{Some scale has significant $\ell_2$ mass.}
    Then by assumption, there exists some $a^*_x$ that is a power of $(1+\eta)$, 
    such that $\norm{2}{x_{\calB_{a^*_x}}}^2 \geq \eps_\infty\cdot n/|B|$. 

    \paragraph{Smallest sample size for which spike appears with probability $\Omega(1/|B|)$.}
    Since entries in $\calB_{a^*_x}$ have squared magnitude $\asymp a^*_x$, 
    we have $|\calB_{a^*_x}| \asymp  \delta\cdot n/(a^*_x\cdot |B|)$. 
    So $\Pr{I\in \calB_{a^*}} \asymp \eps_\infty/(a^*B)$. 
    Set $d^*:=a^*/(1+\eta)$, and let $H:= \{I_1,\dots I_{d^*_x}\}\cap \calB_{a^*_x}$. 
    Then 
    \begin{equation}
    \label{see:spike}
        \Pr{H=\emptyset}~=~ (1-|\calB_{a^*}|/n)^{d^*} = 
        (1-\Omega(\frac{\eps_\infty}{a^*_x \cdot B}))^{d^*_x} 
        ~=~ 1-\Omega(\frac{\eps_\infty d^*}{a^*B}) ~=~ 1-\Omega(\frac{\eps_\infty}{|B|})
    \end{equation}

    \paragraph{$\bbE\max\{1,|\mysmalldot{\xi_{d}}{x}|/\tau\}$ 
    behaves akin to the tail probability of $|\mysmalldot{\xi_{d}}{x}|$ at threshold $\tau$.}
    We next reduce the problem to showing heavy-tailedness  
    of $|\mysmalldot{\xi_{d^*_x}}{x}|$. To this end, 
    set $\beta:= (1-\eta/4)\sqrt{1+\eta} = 1+\eta/4-O(\eta^2)$ and  
    let $q := \Pr{ |\mysmalldot{\xi_{d^*_x}}{x}| ~\geq~ \beta\tau}$. 
    Since $\max\{1,|\mysmalldot{\xi_{d^*_x}}{x}|/\tau\}$ is $1$ with probability 
    $1-q$ and is at least $\beta$ with probability $q$, we have 
    \begin{equation}
    \label{prob:red}
        \Ex{\max\{1,|\mysmalldot{\xi_{d^*_x}}{x}|/\tau\}}
        ~\geq~ (1-q)+\beta\cdot q ~=~ 1+q(\eta/4-O(\eta^2))
    \end{equation}

    \paragraph{$|\mysmalldot{\xi_{d^*_x}}{x}|$ has a heavy tail.}
    It remains to estimate the tail probability of $|\mysmalldot{\xi_{d^*_x}}{x}|$. 
    Recall that $\mysmalldot{\xi_{d^*_x}}{x}$ has identical distribution to 
    $\sqrt{n/d^*_x}\cdot(x_{I_1}^2+\dots +x_{I_{d^*_x}}^2)^{1/2}\cdot g$, where $g$ is a standard gaussian. 
    The key observation here is that a scaling factor $\lambda>1$ of a gaussian 
    causes a disproportionate (power-type) increase in the tail probability, i.e., 
    $\Phi(t/\lambda) = \Phi(t)^{1/\lambda^2}$. 
    
    Moreover, the sample size $d^*_x$ was chosen precisely so that $\sqrt{n/d^*_x}\cdot (x_{I_1}^2+\dots +x_{I_{d^*_x}}^2)^{1/2}$ exceeds $\sqrt{1+\eta}$ with significant probability ($\succeq 1/\log n$). Thus if we set $\tau$ so that the tail probability at 
    threshold $\tau$ is $1/\log^{\Omega(1)}n$, constant fluctuations of 
    $\sqrt{n/d^*_x}\cdot (x_{I_1}^2+\dots +x_{I_{d^*_x}}^2)^{1/2}$ cause a polylogarithmic increase in the tail probability, which is more than enough to 
    offset the $1/\log n$ probability of such a fluctuation occurring. 
    We formalize this below
    \begin{align}
    \label{singlescale:prob}
        &\Pr{ |\mysmalldot{\xi_{d^*_x}}{x}| ~\geq~ \beta\tau} ~=~
        \Pr{\sqrt{n/d^*_x}\cdot (x_{I_1}^2+\dots +x_{I_{d^*_x}}^2)^{1/2}\cdot |g|~\geq~ \beta\tau}
        \nonumber \\
        &\geq \Pr{|g|\geq \beta\tau\sqrt{\frac{d^*_x}{a^*_x}}}\cdot 
        \Pr{x_{I_1}^2+\dots +x_{I_{d^*_x}}^2 \geq \frac{a^*_x}{n}} \nonumber\\
        &= \Phi(\beta\tau/\sqrt{1+\eta})\cdot \Pr{x_{I_1}^2+\dots +x_{I_{d^*_x}}^2 \geq \frac{a^*_x}{n}} 
        \nonumber\\
        &\geq \Phi(\tau(1-\eta/4))\cdot \Pr{H\neq \emptyset} \nonumber\\
        &\succeq \Phi(\tau)^{(1-\eta/4)^2}\cdot \frac{\eps_\infty}{|B|} \nonumber\\
        &\succeq \Phi(\tau)^{1-\eta/2+\eta^2/16}\cdot \frac{\eps_\infty}{|B|} \nonumber\\
        &\succeq \eps_\infty\cdot |B|\cdot \Phi(\tau)^{1-\eta/4+\eta^2/16}
        \qquad\text{assuming $\Phi(\tau)^{\eta/4}\leq 1/|B|^2$}
    \end{align}

    It follows that 
    \begin{equation}
        \Ex{\max\{1,|\mysmalldot{\xi_{d^*_x}}{x}|/\tau\}} \geq 1+\Omega(q\eta)
        \geq t(1+\delta\cdot |B|\cdot \Phi(t)^{1-\eta/4+\eta^2/16})
    \end{equation}

    We are left with estimating the loss in value after averaging over logarithmically many tests. 
    \paragraph{Spike in value persists even after averaging over all tests at $(1+\eta)$-adic scales.}
    We have
    \begin{align}
        &\Ex{\max\{1,|\mysmalldot{\xi}{x}|/\tau\}} \nonumber\\
        &= \frac{1}{|B|}\cdot \Ex{\max\{1,|\mysmalldot{\xi_{d^*_x}}{x}|/\tau\}}+
        \big(1-\frac{1}{|B|}\big)
        \Ex{\max\{1,|\mysmalldot{\xi_{D}}{x}|/\tau\}~|~D\neq d^*_x}\nonumber\\
        &\text{(Law of Total Expectation)} \nonumber\\
        &\geq \frac{1}{|B|}\cdot \Ex{\max\{1,|\mysmalldot{\xi_{d^*_x}}{x}|/\tau\}}+
        \big(1-\frac{1}{|B|}\big)\nonumber\\
        &(\text{since } \max\{1,a\}\geq 1)\nonumber\\
        &\geq \frac{1+\eps_\infty\cdot |B|\cdot \Phi(\tau)^{1-\eta/4+\eta^2/16}}{|B|}+
        \big(1-\frac{1}{|B|}\big)\nonumber\\
        &\text{(by \cref{singlescale:prob})}\nonumber \\
        &\geq 1+\eps_\infty\cdot \Phi(\tau)^{1-\eta/4+\eta^2/16}
        ~~\geq~ 1+\eps_\infty\cdot \Phi(\tau)^{1-\eta/5} \nonumber\\
        &\text{(assuming $\eta/5\leq \eta/4-\eta^2/16$)}
    \end{align}    
    as desired. 
\end{proof}

We defer to \cref{inv:sec} the routine concentration+net argument required to obtain a version of \cref{bdd:test} of polynomial dimension~(\cref{inversion:map}). 
Assuming this, we immediately obtain inapproximability of an affine variant of distortion that is easy to homogeneize later.

\section{Inapproximability of $\ell_2$-radius of an affine slice of $\mathrm{Ball}(\ell_1^m)$}
\label{reduction}
In this section we deduce inapproximability of the $\ell_2$-radius of a low-distortion affine slice of 
$\mathrm{Ball}(\ell_1^m)$, which can be easily homogeneized~(\cref{homogeneization:hardness}) to get hardness of subspace distortion. 
\begin{proposition}[Inapproximability of $\ell_2$-radius of an affine slice of $\mathrm{Ball}(\ell_1^m)$]~\\
\label{aff:sub:hardness}
    There exists a universal constant 
    $\delta\in (0,1)$ such that the following holds. 
    Given an input affine subspace $W\subseteq \R^m$ with 
    the following promised structure: 
    \begin{compactenum}
        \item[(Logarithmic Distortion):] $\Delta_{1\to 2}(W)\preceq \log_{1+\eta}^{3/2}n$
        \item[(Far from Origin in $L_1$):] $\norm{L_1([m])}{y}\geq 1-1/n^{\Omega(1)}$ for all $y\in W$\mcom
    \end{compactenum}
    \smallskip
    
    \noindent
    it is \NP-Hard to distinguish between the following cases: 
    \begin{compactenum}
        \item[(Completeness):] There exists $y\in W\cap \lambda_1\cdot \mathrm{Ball}(L_1([m]))$ satisfying $\norm{L_2([m])}{y}^2\geq 1+(1-o(1))/\tau^2$. 

        \item[(Soundness):] For all $y\in W\cap  (1+\eps_{tol})\cdot\lambda_1\cdot\mathrm{Ball}(L_1([m]))$, we have $\norm{L_2([m])}{y}^2\leq 1+(1+o(1))\delta/\tau^2$. 
    \end{compactenum}
    where $\lambda_1 = (1+1/n^{\Omega(1)})\norm{L_1(\R)}{1+g/\tau}$. 
\end{proposition}
Our starting point will be inapproximability of the $\ell_2$-distance of a (given) subspace from the hypercube (which is closely related to maximizing a PSD quadratic form over the hypercube). This result is proved using a 
local gadget reduction from the PCP theorem (and therefore falls under the local testing framework). 
Closeness to the hypercube is a type of measure of spreadness, and therefore sparse solutions 
are easily dealt with since they have low objective value, which is in stark contrast to $\Delta_{1\to 2}$ where 
sparse solutions have high objective value. 

We use the following result of Briet, Regev and Saket: 
\begin{theorem}[Inapproximability of $\ell_2$-Distance of a Subspace from the Hypercube~\cite{BRS15}]~\\
\label{cube:subspace}
    For any fixed $\eps\in (0,1/10)$, 
    given a subspace $V\subset \R^n$ as input, it is NP-Hard to distinguish between the following cases: 
    \begin{compactenum}
        \item[(Completeness):] $V\cap\{\pm \sfrac{1}{\sqrt{n}}\}^n \neq \emptyset$ 
        
        \item[(Soundness):] $\mathrm{dist}_2(V, \{\pm \sfrac{1}{\sqrt{n}}\}^n)\geq \sqrt{1-2/\pi}-\eps > 1/2$. 
    \end{compactenum}
\end{theorem}

\begin{remark}
    The result of \cite{BRS15} is discussed in a different form as follows: 
    Given a subspace $V$ it is hard to distinguish between the case where 
    $V\cap\{\pm \frac{1}{\sqrt{n}}\}^n \neq \emptyset$ and the case where 
    $\norm{2}{\Pi_V x}\leq \sqrt{2/\pi}+\eps$ for all $x\in \{\pm \frac{1}{\sqrt{n}}\}^n$. To obtain the form above, we simply observe that 
    $\mathrm{dist}_{2}(x,V)^2= \norm{2}{x}^2-\norm{2}{\Pi_V x}^2$. 
\end{remark}

\cref{aff:sub:hardness} is a quick consequence of composing the inversion map \cref{inversion:map} with 
\cref{cube:subspace}. In a nutshell, the inversion map allows one to use $\Delta_{1\to 2}$ to 
test $\ell_2$-proximity to the hypercube.

\subsection{Reduction}
\noindent
\textbf{Input:} Subspace $V\subseteq \R^{n}$, corresponding to an instance of $\mathrm{dist}_2(V,\{\pm\sfrac{1}{\sqrt{n}}\}^n)$. 
\smallskip

\noindent
\textbf{Parameter Setting:}
\begin{compactenum}[-]
    \item Set $\tau := \sqrt{(160/\eta)\cdot \log \log n}$. Note that $\Phi(\tau)\asymp \log^{O(1/\eta)}n$. 
    
    \item Set $\eps_{tol} :=\Phi(\tau)^{1-\eta/10}$.  
    
\end{compactenum}
\smallskip

\noindent
\textbf{Output Instance:}
\begin{align}
\label{affine:subspace:reduction}
    \max &~~\norm{L_2([m])}{y}^{2}
    \quad s.t.
    \nonumber \\
    y \!~&\in T_{inv}(V) 
    &&(\text{Sparsity inversion applied to }V) \nonumber\\
    \norm{L_1([m])}{y} &\leq \norm{L_1(\R)}{1+g/\tau} \cdot (1+1/n^{\Omega(1)})
    &&(\text{Test sparsity of }y)
%    & &&\textrm{and Set-Cover Size})\nonumber\\
%    &Ax=z\cdot\one, &&(\textrm{Exact Coverage Constraints})\\
%    &z=\mysmalldot{x}{\one}/k, &&(\textrm{Homogeneization})
\end{align}

\begin{proof}[\underline{\large Proof of \cref{aff:sub:hardness}}:]~\\~\\

By \cref{inversion:map}, we have the far from origin property in $L_1$. 
Note that $T_{inv}(V)\subseteq \mathrm{Im}(T_{inv})$. So again by \cref{inversion:map} 
we have the logarithmic distortion bound:
\[
    \Delta_{1\to 2}\Big(\bigcup_{\lambda\in \R}\lambda\cdot T_{inv}(V)\Big)
    ~\leq ~~
    \Delta_{1\to 2}\Big(\bigcup_{\lambda\in \R}\lambda\cdot \mathrm{Im}(T_{inv})\Big) ~\preceq ~
    \log_{1+\eta}^{3/2} n \mper
\]

\noindent
\textbf{\large Completeness:} ~\\
Fix any $x\in V\cap \{\pm \sfrac{1}{\sqrt{n}}\}^n$, which exists by the completeness guarantee of \cref{aff:sub:hardness}. 
Let $y:= T_{inv} \, x$. Further note by \cref{inversion:map}, that $\norm{L_1([m])}{y}\leq (1+1/n^{\Omega(1)})\norm{L_1(\R)}{1+g/\tau}$. So $y$ is feasible for \cref{affine:subspace:reduction}, satisfies $\norm{L_2([m])}{y}\geq (1-o(1))(1+1/\tau^2)$. 
\bigskip

\noindent
\textbf{\large Soundness:}~\\
Consider any $y\in T_{inv}(V)$ satisfying $\norm{L_1([m])}{y}\leq (1+\eps_{tol})$,
and let $x:= T_{inv}^\dagger\, y$ (recall $T_{inv}^\dagger$ 
is the pseudoinverse of $T_{inv}$, which coincides with the inverse of 
$T_{inv}$ when restricted to $\mathrm{Im}(T_{inv})$). We know $\norm{L_2([m])}{y}=(1\pm 1/n^{\Omega(1)})(1+\norm{2}{x}^2/\tau^2)$ by the isometry property. 
Fix any sufficiently small universal constant $\eta>0$, and assume for sake of contradiction that $\norm{2}{x}\geq 1-\eta$. Otherwise 
$\norm{L_2([m])}{y}^2\leq (1+o(1))(1+(1-\eta)^2/\tau^2)$, and we're done. 
We will next show that $V$ is close to $\{\pm \sfrac{1}{\sqrt{n}}\}^n$, 
thereby contradicting soundness of \cref{cube:subspace}. 

\paragraph{$x$ is close to continuous cube in $\ell_2$ metric.}
We know $x\in V$. We also know $\norm{L_1([m])}{T_{inv}\,x}\leq (1+\eps_{tol})$, 
and so assuming $\eta/5 <=\eta/4-\eta^2/16$, by \cref{inversion:map}, we must have 
$\mathrm{dist}_2(x, \sqrt{(1+\eta)/n}\cdot \mathrm{Ball}(\ell_\infty^n))\leq \sqrt{\eps_{\infty}}$, 
for $\eps_{\infty}$ satisfying $\eps_{\infty}\cdot\Phi(\tau)^{1-\eta/5} \preceq \eps_{tol}$, which 
implies that $\eps_{\infty}\preceq \Phi(\tau)^{\eta/10} \asymp 1/\log^4 n$. 
Thus 
$\mathrm{dist}_2(x, \sqrt{(1+\eta)/n}\cdot \mathrm{Ball}(\ell_\infty^n))\preceq  1/\log^{2}n$.
Let $\Pi_\infty(x)$ denote the $\ell_2$ projection of $x$ onto $\sqrt{(1+\eta)/n}\cdot \mathrm{Ball}(\ell_\infty^n)$, which must then satisfy $\norm{2}{x-\Pi_\infty(x)}\asymp 1/\log^{2}n$. 
We next show $\Pi_\infty(x)$ is close to $\{\pm \sfrac{1}{\sqrt{n}}\}^n$. 

\paragraph{Rounding $\Pi_\infty(x)$ to $\{\pm \sfrac{1}{\sqrt{n}}\}^n$.}
Observe that $\norm{2}{\Pi_\infty(x)}\geq 1-\eta-o(1)$. 
Set $x^\pm:=\mathrm{sgn}[\Pi_\infty(x)]/\sqrt{n}$ (applied entrywise, and mapped to an arbitrary sign if $(\Pi_\infty(x))_i=0$). We will show that $\Pi_\infty(x)$ is close to $x^\pm$ in $\ell_2$ distance. 
We have  
\begin{equation}
\label{cube:dist}
    \norm{2}{\Pi_\infty(x)-x^\pm}^2 = \sum_i (1/\sqrt{n}-|(\Pi_\infty(x))_i|)^2 = 1+\norm{2}{\Pi_\infty(x)}^2
    -2\cdot \norm{1}{\Pi_\infty(x)}/\sqrt{n}
\end{equation}
On the other hand, by Holder's inequality we have 
$\norm{1}{\Pi_\infty(x)}\geq \norm{2}{\Pi_\infty(x)}^2/\norm{\infty}{\Pi_\infty(x)} \geq \sqrt{n}\cdot \norm{2}{\Pi_\infty(x)}^2/(1+\eta)$. 
Combining this estimate with \cref{cube:dist} and the lower bound on $\norm{2}{\Pi_\infty(x)}$, we obtain $\norm{2}{\Pi_\infty(x)-x^\pm}^2=O(\eta)$. 

By triangle inequality, $\norm{2}{x-x^\pm}=O(\sqrt{\eta})$, which contradicts \cref{cube:subspace}, since $x^\pm\in \{\pm \sfrac{1}{\sqrt{n}}\}^n$. 
Thus $\norm{2}{y}^2\leq (1+o(1))(1+(1-\eta)^2/\tau^2)$ as desired. 
\end{proof}

We are finally ready to prove hardness of $\Delta_{1\to 2}$. The final idea is to homogeneize the instance 
in \cref{aff:sub:hardness} by introducing a new variable $z$ with a large scaling so that all high value solutions 
must set $|z|$ close to $1$. The logarithmic bound on distortion from \cref{inversion:map} is crucial to the success of this reduction. 

\section{Homogeneization}
We reduce from \cref{aff:sub:hardness}. 
We first reduce to the below instance of distortion with a diagonal quadratic objective. It is then easily converted to 
the desired form of $\Delta_{1\to 2}(U)$ for a subspace $U\subseteq \R^{1+m}$. Given an affine subspace $W\subseteq \R^m$, 
we consider the following instance 
\begin{align}
\label{homogeneized}
    &\max~~ z^2+\eps_{hom}\cdot\norm{L_2([m])}{y}^2 ~~~s.t. \nonumber\\
    &y\in z\cdot W, \nonumber\\
    &\|y\|_{L_1([m])}\leq \norm{L_1(\R)}{1+g/\tau} \nonumber \\
    &y\in\R^m,~z\in\R
\end{align}

\noindent
\textbf{Parameter Setting:}
Let $\widetilde{W}:= \cup_{\lambda\in \R} \lambda\cdot W$, which by\cref{inversion:map} 
satisfies the distortion bound $\Delta_{1\to 2}(\widetilde{W})\preceq \log^{3/2}_{1+\eta} n$. 
It suffices to set $\eps_{hom} \asymp \Phi(\tau)^{1-\eta/20}$, so that 
for $n$ sufficiently large, 
\begin{equation}
\label{eps:hom}
    \eps_{hom}\cdot \Delta_{1\to 2}(\widetilde{W})^2 +O(\Phi(\tau)) ~
    \leq  \widetilde{O}(\log^3 n/\log^{(80/\eta)(1-\eta/20))} n) ~\leq~ \Phi(\tau)^{1-\eta/10} ~=~\eps_{tol}\mcom
\end{equation} 
where $\widetilde{O}$ above hides a $\sqrt{\log \log n}$ factor.

\begin{proposition}[Reduction from Affine Distortion to Subspace Distortion]~\\
\label{homogeneization:hardness}
    There is a universal constant $c>1$ such that the following holds. 
    Given a subspace $U\subseteq \R^{1+m}$ as input, it is \NP-Hard to approximate $\Delta_{1\to 2}(U)$ 
    within $1+1/\log^{c}n$. 
\end{proposition}

\begin{proof}
    By completeness of \cref{aff:sub:hardness}, there exists $y\in W$ such that, $\norm{L_1([m])}{y}\leq (1+1/n^{O(1)})\norm{L_1(\R)}{1+g/\tau}\leq 1+C\Phi(\tau)$, and  
    $\norm{L_2([m])}{y}^2\geq (1-o(1))(1+1/\tau^2)$. 
    $(1\oplus y)/(1+1/n^{\Omega(1)})$ is feasible for \cref{homogeneized} and has objective value 
        \[(1-o(1))(1+\eps_{hom}(1+1/\tau^2))\mper\]
    \paragraph{Soundness.}
    Consider any $z,y$ feasible for \cref{homogeneized}. 
    We start by establishing upper and lower estimates on $|z|$. 
    By the far from origin property in \cref{aff:sub:hardness}, we know $|z|\leq (1+1/n^{-\Theta(1)})\norm{L_1([m])}{y} \leq 1+O(\Phi(\tau))$. 

    \paragraph{Small $|z|$ implies low objective value due to logarithmic distortion.}
    On the other hand if $z^2< 1-5\eps_{hom}\cdot \Delta_{1\to 2}(\widetilde{W})^2$, 
    then the objective is at most 
    \[1-5\eps_{hom}\cdot\Delta_{1\to 2}(\widetilde{W})^2 +\eps_{hom}\cdot \norm{L_2([m])}{y}^2 < 
    1-\eps_{hom}\cdot\Delta_{1\to 2}(\widetilde{W})^2\]
    where we used the fact that $\norm{L_1([m])}{y}\leq 1+O(\Phi(\tau))\leq 2$ and $y\in z\cdot W$ for $n$ sufficiently large. 

    \noindent
    So we may assume henceforth that $z^2\in [1-O(\eps_{hom}\cdot \Delta_{1\to 2}(\widetilde{W})^2),\,1+O(\Phi(\tau))]$. 

    \paragraph{Rounding $z\oplus y$ to affine subspace $W$.}
    It follows that $y/z$ satisfies $\norm{L_1([m])}{y/z}= 1+O(\eps_{hom}\cdot \Delta_{1\to 2}(\widetilde{W})^2)+O(\Phi(\tau))
    \leq 1+\eps_{tol}\mcom$ for $n$ sufficiently large, where the final step 
    follows from \cref{eps:hom}. 
    By soundness of \cref{aff:sub:hardness}, $\norm{L_2([m])}{y/z}^2\leq (1+1/n^{\Omega(1)})(1+\delta/\tau^2)$ for some fixed $\delta<1$, which implies $\norm{L_2([m])}{y}^2\leq (1+O(\Phi(\tau)))(1+\delta/\tau^2)=(1+\delta(1+o(1))/\tau^2)$, since $|z|\leq 1+O(\Phi(\tau))$. 
    Thus objective is at most 
    \[
        1+O(\Phi(\tau))+\eps_{hom}(1+\delta(1+o(1))/\tau^2)= (1+o(1))(1+\eps_{hom}(1+\delta)/\tau^2) 
    \]  
    We obtain a multiplicative gap of $1+\Omega(\eps_{hom}/\tau^2)=1+1/\log^{O(1)}n$ as desired.
    \bigskip 

    \paragraph{Final Subspace Distortion Instance.} It is convenient to switch back to counting norms for the remainder of this proof. 
    We first observe that replacing the constraint $\|y\|_{L_1([m])}\leq \norm{L_1(\R)}{1+g/\tau} $ 
    by $|z|/m^2+\|y\|_{1}/m\leq \norm{L_1(\R)}{1+g/\tau}$ preserves the gap since the first term is negligible for $m$ is sufficiently large 
    and the $\|y\|_{1}/m$ term is unchanged. The objective of \cref{homogeneized} can be written as $z^2+\eps_{hom}\cdot \norm{2}{y}^2/m$. 
    Then we use a standard duplication trick -- replace $y$ by $k$ concatenated copies of $(y/k)$. This doesn't change 
    the $\ell_1$ norm for any $y$, and on the other hand, $\norm{2}{(y/k)^{\oplus k}}^2 = \norm{2}{y}^2/k$.
    We then set $k$ precisely so that $\eps_{hom}/m^3 = 1/k$ (we may choose $\eps_{hom},k$ without changing their asymptotics so that there are no integrality issues). Thus for this $k$, we define the subspace 
    \[
        U:= \{(z/m)\oplus (y/k)^{\oplus k}~:~y\in z\cdot W,~z\in\R\}
    \]
    and moreover we have 
    \[
        \norm{2}{(z/m)\oplus (y/k)^{\oplus k}}^2 = z^2/m^2+\norm{2}{y}^2/k = (1/m^2)(z^2+\eps_{hom}\norm{2}{y}^2/m)
    \]
    and 
    \[
        \norm{1}{(z/m)\oplus (y/k)^{\oplus k}} = |z|/m+ \norm{1}{y}\mper
    \]
    Thus $\Delta_{1\to 2}(U)$, upto a scaling factor, a $1+1/m^{\Omega(1)}$ approximation to \cref{homogeneized}, and we are done. 
\end{proof}

\iffalse
\[
        \norm{L_1([m])}{y}\leq (1+\eps_{hom}-(1-2\eps_{hom}\cdot \Delta_{1\to 2}(\widetilde{W})^2))/\eps_{hom} \quad 
        \Rightarrow \norm{L_1([m])}{y}\leq 1+2\eps_{hom}^3\cdot \Delta_{1\to 2}(\widetilde{W})^2\mper
    \]

Suppose $\Delta_{1\to 2}(b+V)\leq 1-\eps_{gap}$, and 
    suppose $\eps_{gap}\gg \Delta_{1\to 2}(V)\cdot \eps_{hom}$. Then 
    \[
        \Delta_{1\to 2}(V_{fin}) \leq 1-\eps_{gap}+\Delta_{1\to 2}(V)\cdot \eps_{hom}
    \]
\fi

\subsection*{Amplification by Tensoring}
For subspaces $U\subseteq \R^m,~V\subseteq \R^n$, the tensor product $U\otimes V$ is defined as $\mathrm{Span}\{u\otimes v : u\in U,~v\in v\}$ and is equal to the dual description 
of the set of $m\times n$ matrices $M$ such that all columns of $M$ lie in $U$ and all rows of $M$ lie in $V$. 
We amplify the gap using the usual notion of tensoring. The below result is likely folklore and 
we do not know the earliest reference. 
\begin{proposition}[Distortion Tensorizes~(e.g. Theorem 4.11 in \cite{BGGLT23})]~\\
\label{tensor}
For any $1\leq p\leq q\leq \infty$, and any subspaces $U\subseteq \R^m,~V\subseteq \R^n$, ~
$
    \Delta_{p\to q}(U\otimes V) = \Delta_{p\to q}(U)\cdot \Delta_{p\to q}(V)
$. 
\end{proposition}
\begin{remark}
    The proof in \cite{BGGLT23} is for the operator version, but it can be adapted essentially 
    verbatim to the subspace version. 
\end{remark}

Composing \cref{homogeneization:hardness}, and \cref{tensor} yields almost polynomial hardness for Distortion, and this can be extended to $\Delta_{p\to 2}$ for $p=1+1/\log^\eps n$, using the well known fact that $\norm{1}{x}/\norm{p}{x}\leq n^{1-1/p}$. 
\main*

\section{Polynomial Dimension Sparsity Inversion Map via Net Argument}
\label{inv:sec}
In this section we show that averaging polynomially many samples suffices, to approximate the map 
$(z\oplus x)\mapsto \Ex{|z+\mysmalldot{\xi}{x}/\tau|}$ uniformly over $\R^{1+ n}$. 
We also obtain other similar properties enroute (namely near-isometry and logarithmic distortion) that are needed for our reduction. 
Thus in this section we complete the proof of the following:
\begin{theorem}[Near-Isometric Affine Map, Inverting Sparsity on $\SSS^{n-1}$]~\\
\label{inversion:map}
    There is an affine linear map 
    $T_{inv}:\R^n\to \R^m$ computable in 
    randomized $\mathrm{poly}(m,n)$ time, such that the following 
    holds. Fix any $\eps_\infty,\eta\in (0,1)$ such that $\eta/5\leq \eta/4-\eta^2/16$. Then with probability $1-e^{-\Omega(n)}$, $T_{inv}$ has the following properties 
    \begin{enumerate}
        \item[(Near Isometry):] $\forall x\in\R^n,~ \norm{2}{T_{inv} \, x}=
        (1\pm 1/n^{\Omega(1)})(1+\norm{2}{x}^2/\tau^2)^{1/2}$. 
        
        \item[($\{\pm \sfrac{1}{\sqrt{n}}\}^n$ has a Sparse Image):] For any 
        $x\in \{\pm \sfrac{1}{\sqrt{n}}\}^n$, ~~ 
        \[
            \norm{L_1([m])}{T_{inv}\, x}\leq (1+1/n^{\Omega(1)})\cdot\norm{L_1(\R)}{1+g/\tau} \leq 
            1+O(\Phi(\tau))\mper 
        \]

        \item[(Sparse Points Have a Spread Image):] For any $x$ satisfying 
        $\mathrm{dist}_2(x,\sqrt{(1+\eta)/n}\cdot \mathrm{Ball}(\ell^n_\infty))\geq \sqrt{\eps_\infty}$,  
        we~have 
        \[
            \norm{L_1([m])}{T_{inv}\, x} \geq 1+\Omega(\eps_\infty\cdot \Phi(\tau)^{1-\eta/5})
        \] 

        \item[(Logarithmic Distortion):] 
        \[
            \Delta_{1\to 2}\Big(\bigcup_{\lambda\in\R}
            \lambda\cdot \mathrm{Im}(T_{inv})\Big)
            ~\preceq ~
            \log_{1+\eta}^{3/2} n
        \]

        \item[(Far from Origin in $L_1$):] 
        $\norm{L_1([m])}{y}\geq 1-1/n^{\Omega(1)}$ for all $y\in \mathrm{Im}(T_{inv})$\mcom
    \end{enumerate}
    where $\Phi(\tau)= \widetilde{O}( 1/\log^{80/\eta}n)$. 
\end{theorem}

\cref{inversion:map} follows from combining \cref{bdd:test} with the ensuing ``concentration + net union bound'' argument which is by now quite routine in high dimensional geometry.

\subsection{Empirical Estimator for $\|1+\mysmalldot{\xi}{x/\tau}\|_{L_1(\R)}$}
    Define $\widehat{\Xi}\in \R^{m\times n}$ 
    be a random matrix whose rows are i.i.d. copies of $\xi/\tau$.
    Define the (random) affine linear map $T_{inv}: \R^{n}\to \R^m$ 
    as \[T_{inv}\,x := \one+\widehat{\Xi} \,x \mper\] 
    Also define the (random) linear map $\widetilde{T}: \R^{1+n}\to \R^m$ 
    as $\widetilde{T}(z\oplus x) := z\cdot T_{inv}\,(x/z)$.

\subsection{Proof of Concentration for \cref{inversion:map}}
\label{conc}
We begin by concluding the near-isometry property from off-the-shelf results. 
\paragraph{Near-Isometry.}
    For any $d\in \N$, $\xi_d$ is isotropic, namely $\Ex{\xi_d\xi_d^T} = n\cdot \sum_{j,\ell} \Ex{(g_jg_\ell/d)e_{I_j}e_{I_\ell}^T} = \id$. 
Thus the same holds for $\xi$: $\Ex{\xi\xi^T}=\id$. 
It follows that for any $z\in \R$, $x\in\R^n$, 
\begin{equation}
\label{isotropy}
    \Ex{(z+\mysmalldot{\xi}{x}/\tau)^2} = z^2+\norm{2}{x}^2/\tau^2
\end{equation}
It follows from Theorem 4.6.1 in \cite{vershynin_2018} that for $m\geq n^6$, with probability $1-e^{-\Omega(n)}$, 
we have for all $(z\oplus x)\in\R^{1+n}$, 
\[
    \norm{L_2[m]}{\widetilde{T}(z\oplus x)} = (1\pm 1/n^{\Omega(1)})\cdot \norm{L_2(\R)}{z+\mysmalldot{\xi}{x}/\tau}\mper
\]

\noindent
We proceed to prove the desired uniform approximation for the empirical estimator. 

\noindent
    \textbf{Error Estimate for Rounding to Net Point.}
    Let $\mathcal{N}\subset [-1,1]\times \mathrm{Ball}(\ell_2^{n})$ be an $\eps_{net}$-net of $[-1,1]\times \mathrm{Ball}(\ell_2^{n})$,  according to the norm $\|z\oplus x\|:=\max\{|z|,\norm{2}{x}\}$ (which is also the minkowski functional of $[-1,1]\times \mathrm{Ball}(\ell_2^{n})$), of size at most $(10/\eps_{net})^{n}$ 
    (which exists by a standard volume ratio argument). Let $(z\oplus x)\in [-1,1]\times \mathrm{Ball}(\ell_2^{n})$ attain the maximum $\lambda^*$ of 
    $\norm{L_1([m])}{\widetilde{T} (z\oplus x)}$ over $[-1,1]\times \mathrm{Ball}(\ell_2^{n})$. 
    There exists $(z'\oplus x')\in \mathcal{N}$ of $\ell_2$ distance at most $\eps_{net}$ from $(z\oplus x)$. 
    Then by triangle inequality, 
    \[  
        \norm{L_1([m])}{\widetilde{T} (z'\oplus x')} \geq 
        \norm{L_1([m])}{\widetilde{T} (z\oplus x)}-\lambda^*\cdot \max\{|z-z'|,\norm{2}{x-x'}\} \geq 
        \lambda^*(1- O(\eps_{net}))
    \]
    \iffalse
    \[  
        \norm{1}{T_{inv} v} \geq 
        \norm{1}{T_{inv} u}-\norm{2}{1}{T_{inv}}\norm{2}{u-v} \geq 
        \norm{2}{1}{T_{inv}}(1- \eps_{net})
    \]
    \fi
    
    Therefore \[\max_{(z'\oplus x')\in \mathcal{N}}\norm{L_1([m])}{\widetilde{T} (z'\oplus x')} \geq 
    (1-O(\eps_{net}))\lambda^*\mper\]
    \bigskip 
    
    \noindent
    \textbf{Concentration.} 
    Let $\norm{\Psi_2}{\cdot}$ denote the subgaussian norm, i.e. the minimum scaling $\lambda$ of a random variable $X\in\R$ 
    such that $\inf\{t>0~:~\Ex{e^{X^2/t^2}\leq 2}\}$
    We have
    \[
        |z+\sqrt{n/d}\cdot (x_{I_1}^2+\dots +x_{I_d}^2)^{1/2}g/\tau|
        \leq 
        |z|+\sqrt{n}\cdot\norm{2}{x}\cdot |g|/\tau
        \leq |z|+\sqrt{n}\cdot |g|/\tau
    \]
    From this we obtain the following estimate
    \begin{equation}
    \norm{\Psi_2}{z+\mysmalldot{\xi}{x}/\tau}
    ~\leq~ |z|+\sqrt{n}\norm{\Psi_2}{g}/\tau ~\leq~ |z|+\sqrt{n}/\tau
    \end{equation}
    Since for any $x\in \mathrm{Ball}(\ell_2^{n})$ and $z\in [-1,1]$, 
    $\norm{L_1([m])}{\widetilde{T} (z\oplus x)}$ is the average of $m$ independent  subgaussians with norm $\preceq \sqrt{n}$, we have by the subgaussian concentration inequality that 
    \[
        \Pr{\left|\norm{L_1([m])}{\widetilde{T} (z\oplus x)}-
        \norm{L_1(\R)}{z+\mysmalldot{\xi}{x}/\tau}\right|
        > \Delta/m}\leq \frac{1}{e^{\Omega(\Delta^2/(mn))}}
    \]
    Setting $\Delta:=\eps_{net}\cdot m$, we obtain that so long as 
    $m\geq C'\cdot n^2\log (1/\eps_{net})/\eps_{net}^2$ for a sufficiently large constant $C'$, the following holds with probability at least $1-e^{-\Omega(n)}$, 
    \[
        \forall (z\oplus x)\in \mathcal{N},~~\left|
        \norm{1}{\widetilde{T} (z\oplus x)}-\norm{L_1(\R)}{z+\mysmalldot{\xi}{x}/\tau}\right|\leq \eps_{net}
    \]

    \paragraph{Error Estimation.}
    We have 
    \[
        \norm{L_1(\R)}{z-z'+\mysmalldot{\xi}{x-x'}/\tau}
        \leq \norm{L_2(\R)}{z-z'+\mysmalldot{\xi}{x-x'}/\tau}
        \leq (z-z')^2+\norm{2}{x-x'}/\tau^2
    \]
    \iffalse
    But we also know 
    \[
        |\mathrm{Test}_\infty(u)-\mathrm{Test}_\infty(v)| 
        \leq \mathrm{Test}_\infty(u-v)
        \leq \norm{2}{u-v}
    \]
    \fi
    It follows that for all $(z\oplus x)\in [-1,1]\times\mathrm{Ball}(\ell_2^{n})$, 
    \begin{equation}
    \label{conc:l1}
        \left|\norm{L_1([m])}{\widetilde{T} (z\oplus x)}-
        \norm{L_1(\R)}{z+\mysmalldot{\xi}{x}/\tau}\right|
        \leq 2\eps_{net}
    \end{equation}
    We will set $\eps_{net}:=1/n^2$. It remains to argue that the expectation is much larger than $\eps_{net}$. 
    
    \paragraph{Upper and Lower Bounds on Expectation.}
    Fix any $x\in \mathrm{Ball}(\ell_2^{n})$ and $z\in [-1,1]$. 
    Then 
    \begin{equation}
    \label{up:est}
        \Ex{(z+\mysmalldot{\xi}{x}/\tau)^2}^{1/2}
        = 
        (z^2+\norm{2}{x}^2/\tau^2)^{1/2}
        \leq \sqrt{2}\cdot \max\{|z|,\norm{2}{x}/\tau\}
    \end{equation}
    and 
    \begin{align}
        &\Ex{|z+\mysmalldot{\xi}{x}/\tau|}
        ~=~
        \Ex{\max\{|z|,|\mysmalldot{\xi}{x}|/\tau\}} 
        ~\geq ~
        \frac{1}{\tau}\Ex{|\mysmalldot{\xi}{x}|}
        ~\geq~  
        \frac{1}{\tau\cdot |B|}\Ex{|\mysmalldot{\xi_{\overline{n}}}{x}|} \nonumber \\
        &= \frac{\sqrt{2}}{\sqrt{\pi}\tau\cdot |B|} 
        \Ex{\sqrt{n/\bar{n}}\cdot (x_{I_1}^2+\dots +x_{I_{\overline{n}}}^2)^{1/2}} 
        \succeq \frac{1}{\tau\cdot \log^{3/2}_{1+\eta} n}
        \norm{2}{x}\cdot \Pr{\{I_1,\dots ,I_{\bar{n}}\}\supset [n]}
    \end{align}
    where $\Pr{\{I_1,\dots ,I_{\bar{n}}\}\supset \![n]\,}$ can be easily lower bounded by considering the complement event and taking union bound over all events of the form ``element $i$ is not contained in $\{I_1,\dots ,I_{\bar{n}}\}$'', which occurs with probability at most $(1-1/n)^{\bar{n}}\leq e^{-\bar{n}/n}$. Thus the probability is at least $1-n\cdot e^{-\bar{n}/n} \geq 1-1/n$. Thus 
    \begin{equation}
    \label{exp:lb}
        \Ex{|z+\mysmalldot{\xi}{x}/\tau|}\geq \max\{|z|,\norm{2}{x}/(\tau\cdot\log^{3/2}_{1+\eta} n)\}
        \geq 
        \max\{|z|,\norm{2}{x}/\tau\}/\log^{3/2}_{1+\eta} n
    \end{equation}

    \paragraph{Uniform Approximation.}
    Combining \cref{exp:lb} and \cref{conc:l1}, whenever $\max\{|z|,\norm{2}{x}\}= 1$ and $n$ is sufficiently large, 
    we have $\norm{L_1([m])}{\widetilde{T} (z\oplus x)}=(1\pm 2\eps_{net}). 
        \norm{L_1(\R)}{z+\mysmalldot{\xi}{x}/\tau}$. However both the LHS and RHS scale by $\lambda$ if we scale $(z\oplus x)$ by any $\lambda>0$. Since any $(z'\oplus x')\in\R^{1+n}$ is a scaling of $(z\oplus x)$ satisfying $\max\{|z|,\norm{2}{x}\}= 1$,we obtain 
        the same multiplicative distortion for all of $\R^{1+n}$. Thus  
        \begin{equation}
        \label{uniform:l1}
            \norm{L_1([m])}{\widetilde{T} (z\oplus x)}=(1\pm 2\eps_{net}). 
            \norm{L_1(\R)}{z+\mysmalldot{\xi}{x}/\tau}\mcom \quad\forall~ (z\oplus x)\in\R^{1+n}
        \end{equation}
    as desired. 
    The far from origin property in $L_1$ also follows from the above discussion and the fact that $\norm{L_1(\R)}{1+\mysmalldot{\xi}{x}/\tau}\geq 1$. 
    
    \paragraph{Distortion.}
    We may also immediately conclude the desired distortion property. 
    Combining \cref{exp:lb} and \cref{up:est}, we get 
    \[
        \Ex{|z+\mysmalldot{\xi}{x}/\tau|^2}^{1/2}/\Ex{|z+\mysmalldot{\xi}{x}/\tau|} \preceq \log_{1+\eta}^{3/2} n\mper
    \]
    It follows from the near-isometry property of $\widetilde{T}$ and \cref{uniform:l1} that 
    \[
        \frac{\norm{L_2([m])}{\widetilde{T}}}{\norm{L_1([m])}{\widetilde{T}}} \preceq \log_{1+\eta}^{3/2} n \mper
    \]
    This completes the proof of \cref{inversion:map}.

\section*{Acknowledgements}
The author is very grateful to an anonymous reviewer for CCC 2026 who flagged an oversight in a previous version of this work. In the pervious manuscript we were missing an important idea -- sampling at all dyadic scales, which took many more months to conceive of. 

The author has had a long journey pursuing the questions in this paper. I wish to thank several individuals that have played a very significant but intangible role in the development of this work over
the years. Madhur Tulsiani, who for instance made the connection to uncertainty principles; 
Euiwoong Lee who used that to show hardness of $4\to 8$ norm and who years later made the very fruitful suggestion of exploring the coding/lattice hardness literature for ideas; {Venkat Guruswami}, for suggesting a connection to convex geometry and embedding, which ended up being the right general area for discovering the phenomena governing these problems; \\
{Assaf Naor}, who identified and greatly deepened my understanding of the relevant parts of convex geometry in a short amount of time (type-$2$, factorization norms, $p$-stable embeddings, etc.); 
{Mrinalkanti~Ghosh}, who pointed me towards the right parts of complex analysis when working on the algorithmic side of these questions years ago.

\bibliographystyle{alpha}
\bibliography{refs}

%\section{Old Overlap}

\end{document}